\documentclass[aps,preprint,showpacs,nofootinbib]{revtex4}
\usepackage{amsfonts}
\usepackage{amsmath}
\usepackage{amssymb}
\usepackage{relsize}
\usepackage{graphicx}
\usepackage{pdfpages}
\usepackage{subfigure}
\usepackage{float}
\usepackage[caption = false]{subfig}
\usepackage{color}
\usepackage{multirow}
\newcommand{\sx}[1]{\scriptsize{#1}}
\newcommand{\ft}[1]{\footnotesize{#1}}
\def\correspondingauthor{\footnote{Corresponding author. E-mail address: liverts@phys.huji.ac.il}}
\begin{document}
\bibliographystyle{plainnat}
\title{Accurate exponential representations for the ground states of the collinear two-electron atomic systems}

\author{Evgeny Z. Liverts\correspondingauthor{}}
\author{Evgeny Z. Liverts}
\affiliation{Racah Institute of Physics, The Hebrew University, Jerusalem 91904, Israel}

\author{Nir Barnea}
\affiliation{Racah Institute of Physics, The Hebrew University, Jerusalem 91904, Israel}

\begin{abstract}
In the framework of the study of helium-like atomic systems possessing the \emph{collinear} configuration, we propose a simple method for computing compact but very accurate wave functions describing the relevant $S$-state.
It is worth noting that the considered states include the well-known states of the \emph{electron-nucleus} and \emph{electron-electron} coalescences as a particular case.
The simplicity and compactness imply that the considered wave functions represent a linear combinations of few single exponentials. We have calculated such model wave functions for the ground state of helium and the two-electron ions with nucleus charge $1 \leq Z \leq 5$.
The parameters and the accompanying characteristics of these functions are presented in tables for number of exponential from 3 to 6. The accuracy of the resulting wave functions are confirmed graphically.
The specific properties of the relevant codes by Wolfram Mathematica are discussed.
An example of application of the compact wave functions under consideration is reported.

\end{abstract}

\pacs{ }

\maketitle

\section{Introduction}\label{S0}

In this paper we present a technique for building compact and simple wave functions of high accuracy, describing two-electron atomic systems
such as H$^-$, He, Li$^+$, Be$^{2+}$ and B$^{3+}$
with the \emph{collinear} arrangement of the particles \cite{LEZ3}.
The study of mechanism of double photoionization of the helium-like atomic systems by high energy photons \cite{AMY1, AMY2} can serve as an example of possible application (see the details in the next Section).

Methods enabling us to calculate the relevant wave function (WF)
and the corresponding non-relativistic energy differ from each other by the calculation technique, spatial variables and basis sets.
It is well-known that the $S$-state WF,  $\Psi(r_1,r_2,r_{12})$ is a function of three variables: the distances $r_1\equiv|\textbf{r}_1|$ and $r_2\equiv|\textbf{r}_2|$ between the nucleus and electrons, and the interelectron distance $r_{12}\equiv|\textbf{r}_1-\textbf{r}_2|$, where
$\textbf{r}_1$ and $\textbf{r}_2$ represent radius-vectors of the electrons.
We shall pay special attention to the bases that differ from each other both in the kind of the basis functions and in its number (basis size).
The Hartree atomic units are used throughout the paper.

It would be useful to give some examples of basis sets intended for describing the relevant $S$ states.
The correlation function hyperspherical harmonic method (CFHHM)  \cite{HM1,HM2} employs the basis representing the product of the hyperspherical harmonic (HH) as an angular part, and the \emph{numerical} radial part. The corresponding basis size $N$ equals (as a rule) $625$.
The Pekeris-like method (PLM) \cite{PEK,LEZ1,LEZ2} is used intensively in the current work.
The basis size of the PLM under consideration is $N=1729$ (for the number of shells $\Omega=25$), and the basis functions can be finally reduced to the form $\exp(\alpha r_1+\beta r_2+\gamma r_{12})r_1^lr_2^m r_{12}^n$,
where $\alpha,\beta$ and $\gamma$ are the real constants and $l,m,n$ are non-negative integers.
Hylleraas \cite{HYL} (see also \cite{CHAN,KIN}) was the first who employed the same basis but with $\gamma=0$.
The authors of Ref.\cite{FRE} have performed variational calculations on the helium isoelectronic sequence using modification of the basis set that employed by Frankowski and Pekeris \cite{FRA}. They managed to get very accurate results using the reduced basis of the size $N=230$.
The variational basis functions of the type $\exp(\tilde{\alpha} r_1+\tilde{\beta} r_2+\tilde{\gamma} r_{12})$ with complex exponents were used in the works of Korobov \cite{KOR}  ($N=1400-2200$) and Frolov \cite{FR1} for $N=600-2700$ (see also references therein).
Application of the Gaussian bases of the size $N>100$ can be found in the book \cite{SUZ}.
The reviews on the helium-like atomic system and the methods of their calculations can be found, e.g. in the handbook \cite{DRK}.

In this paper we propose a simple method of calculation of the compact but very accurate WFs describing the two-electron atom/ion with collinear configuration.
The results and example of application of the relevant technique are presented in the next sections.

\section{Calculation technique}\label{S1}

The simplicity of the WFs under consideration implies that the form
\begin{equation}\label{1}
f_N(r)=\sum_{k=1}^{N}C_k\exp(-b_k r)
\end{equation}
represents the sum of a few single exponentials, whereas the compactness means that their number $3\leq N\leq 6$ in Eq.(\ref{1}),
unlike the basis sizes mentioned in Introduction.
The relevant accuracy will be discussed later.
It is seen that the RHS of Eq.(\ref{1}) includes $N$ linear parameters $C_k$ and $N$ nonlinear parameters $b_k$ with $k=1,2,...N$.

The \emph{collinear} arrangement of the particles consisting of the nucleus and two electrons can be described by a single scalar parameter $\lambda$ as follows \cite{LEZ3}:
\begin{equation}\label{2}
r_1=r,~~~~~~~~r_2=|\lambda|r,~~~~~~~~r_{12}=(1-\lambda)r,
\end{equation}
where $\lambda \in [-1,1]$, and $r$ is the distance between the nucleus and the electron most distant from it.
Clearly $\lambda=0$ corresponds to the \emph{electron-nucleus} coalescence, and $\lambda=1$ to the \emph{electron-electron} coalescence.
The boundary value $\lambda=-1$ corresponds to the collinear \textbf{e-n-e} configuration with the same distances of both electrons from the nucleus.
In general, $0 < \lambda \leq 1$ corresponds to the collinear arrangement of
the form \textbf{n-e-e} where both electrons
are on the same side of the nucleus.
Accordingly, $-1\leq\lambda < 0$ corresponds to the collinear arrangement
of the form \textbf{e-n-e} where the electrons are on the opposite sides of
the nucleus.
The absolute value $|\lambda|$ measures the ratio of the distances of the electrons from the nucleus.

Thus, for the particles with collinear arrangement we can introduce the \emph{collinear} WF of the form
\begin{equation}\label{3}
\Phi(r,\lambda)\equiv \Psi(r,|\lambda|r,(1-\lambda)r)/\Psi(0,0,0).
\end{equation}
It should be emphasized that, e.g., the PLM WF with \emph{collinear} configuration reduces to the form
\begin{equation}\label{4}
\Phi_{PLM}(r,\lambda)=\exp\left(-\delta_\lambda r\right)\sum_{p=0}^\Omega c_p(\lambda)r^p,
\end{equation}
where $\Omega=25$ for the current (standard) consideration, as it was mentioned earlier.

We can give an example of the physical problem where the collinear WF of the form (\ref{4}) cannot be applied, but the quite accurate WF of the form (\ref{1}) is required instead.
In Refs. \cite{AMY1, AMY2} the mechanism of photoionization in the two-electron atoms is investigated. Calculations of various differential characteristics (cross sections) of ionization are based on computation of the triple integral of the form
\begin{equation} \label{5}
\int d^{3}\mathbf{r}~e^{i\mathbf{q}\mathbf{r}}~
_{1}F_{1}(i\xi _{1},1,ip_{1}r-i\mathbf{p}_1\mathbf{r})_{1}F_{1}(i\xi _{2},1,ip_{2}r-i\mathbf{p}_2\mathbf{r})
\Phi(r,1),
\end{equation}
where $\mathbf{p}_j$ ($j=1,2$) are the momenta of photoelectrons, $\mathbf{q}$ is the recoil momentum, $\xi_j=Z/p_j$, $i$ is the imaginary unit, and $_{1}F_{1}(...)$ is the confluent hypergeometric function of the first kind.
The most important for our consideration is the fact that integral (\ref{5}) contains the \emph{collinear} WF $\Phi(r,1)$ describing the case of the electron-electron coalescence ($\lambda=1$) in the helium-like atom/ion with the nucleus charge $Z$.
It is clear that the numerical computation of the triple integral (\ref{5}) is not impossible but
rather difficult problem especially for building the relevant graphs.
Fortunately, already at 1954 \cite{NORD} the explicit expression for the triple integral which is very close to integral (\ref{5}) was derived.
In fact, integral (\ref{5}) can be calculated by simple differentiation (with respect to a parameter) of the explicit form for the integral mentioned above, but only under condition that the WF, $\Phi(r,1)$ is represented by a single exponential of the form $\exp(-b r)$ (with positive parameter $b$, of course).

According to the Fock expansion \cite{FOCK} (see also \cite{AB1, LEZ4}), we have:
\begin{equation}\label{6}
\Psi(r_1,r_2,r_{12})/\Psi(0,0,0)\underset{R\rightarrow 0}{=}1-Z(r_1+r_2)+\frac{1}{2}r_{12}-
Z\left(\frac{\pi-2}{3\pi}\right)\left(R^2-r_{12}^2\right)\ln R+O(R^2),
\end{equation}
where $R=(r_1^2+r_2^2)^{1/2}$ is the hyperspherical radius.
Using Eq.(\ref{6}) and the collinear conditions (\ref{2}) we obtain the Fock expansion for the collinear WF in the form:
\begin{equation}\label{7}
\Phi(r,\lambda)\underset{r\rightarrow 0}{=}1+\eta_\lambda r
+\zeta_\lambda r^2\ln r+\xi_\lambda r^2+...
\end{equation}
where
\begin{equation}\label{7a}
\eta_\lambda=-Z \left(1+|\lambda|\right)+\frac{1-\lambda}{2},
\end{equation}
\begin{equation}\label{7b}
\zeta_\lambda=-\frac{2Z\lambda(\pi-2)}{3\pi},
\end{equation}
and the general form of the coefficient $\xi_\lambda$ being rather complicated will be discussed later.
The necessity of the equivalent behavior of the model WF, (\ref{1}) and the variational WF, $\Phi(r,\lambda)$ near the nucleus ($r\rightarrow 0$) results in the following two coupled equations for $2N$ parameters
$\textbf{C}_N\equiv \left\{C_1,C_2,...,C_N\right\}$ and
$\textbf{b}_N\equiv\left\{b_1,b_2,...,b_N\right\} $ of the model WF:
\begin{equation}\label{8}
\sum_{k=1}^N C_k=1,
\end{equation}
\begin{equation}\label{9}
\sum_{k=1}^N C_k b_k=Z(1+|\lambda|)+\frac{\lambda-1}{2}.
\end{equation}
Eq.(\ref{8}) follows from the condition $\Phi(0,\lambda)=1$, whereas Eq.(\ref{9}) is obtained by equating the linear (in $r$) coefficients of the power series expansion of the model WF (\ref{1}) and the Fock expansion (\ref{7}).

As it was mentioned above, to obtain the fully defined model WF of the form (\ref{1}) one needs to determine $2N$ coefficients. To solve the problem with given equations (\ref{8}) and (\ref{9}), we need to find extra $2(N-1)$ coupled equations for parameters of the exponential form (\ref{1}).
To this end we propose to use the definite integral properties of the collinear WF (\ref{3}).

A  number of numerical results presenting expectation values of Dirac-delta functions $\left\langle\delta(\textbf{r}_1) \right\rangle$, $\left\langle\delta(\textbf{r}_{12})\right\rangle\equiv\left\langle  \delta(\textbf{r}_1-\textbf{r}_2)\right\rangle$ and
$\left\langle\delta(\textbf{r}_1)\delta(\textbf{r}_2) \right\rangle$
for the helium-like atoms can be found in the proper scientific literature (see, e.g., \cite{DRK, FR1, FR2} and references therein).
It was shown \cite{LEZ3} that expectation values mentioned above represent the particular cases of the more general expectation value
\begin{equation}\label{10}
\left\langle\delta\left(\textbf{r}_1-\lambda \textbf{r}_2\right)\right\rangle=
4\pi \left\langle\delta(\textbf{r}_1)\delta(\textbf{r}_2) \right\rangle\int_0^\infty \left|\Phi\left(r,\lambda\right)\right|^2r^2 dr,
\end{equation}
where
\begin{equation}\label{11}
\left\langle\delta(\textbf{r}_1)\delta(\textbf{r}_2) \right\rangle=
\Psi^2(0,0,0)/\int \psi^2(\textbf{r}_1,\textbf{r}_2)d^3\textbf{r}_1 d^3\textbf{r}_2
\end{equation}
is a square of the normalized WF taken at the nucleus.
It is seen that expectation value (\ref{10}) is fully defined by the \emph{collinear} WF, $\Phi(r,\lambda)$.

We propose to use the integrals of the form
\begin{equation}\label{12}
S_n=\int_0^\infty \left|\Phi\left(r,\lambda\right)\right|^2r^n dr,~~~~~~~(n=0,1,2,...)
\end{equation}
for deriving $2(N-1)$ extra coupled equations required, in its turn, for determining $2N$ coefficients defining the model WF, (\ref{1}).
Replacing $\Phi(r,\lambda)$ in the RHS of Eq.(\ref{12}) by the model WF (\ref{1}) and using the closed form of the corresponding integral, one obtains $n$ equation of the form
\begin{equation}\label{13}
S_n=n!\sum_{j=1}^N\sum_{k=1}^N \frac{C_j C_k}{(b_j+b_k)^n},
\end{equation}
where, in fact, $\textbf{C}_N\equiv \textbf{C}_N(Z,\lambda)$, $\textbf{b}_N\equiv \textbf{b}_N(Z,\lambda)$ are the coefficients we are requested, whereas the integrals $S_n \equiv S_n(Z,\lambda)$ can be computed using, for example, the PLM WFs according to definition (\ref{12}).
The technique proposed, in fact, represents a variant of the "Method of Moments" (see, e.g., \cite{MMT}) supplemented by the boundary conditions (\ref{8}) and (\ref{9}).

The problem is that it is necessary to select a set (sample) of integers $\left\{n_1,n_2,...n_{2(N-1)}\right\}$
describing the integrals (\ref{12}), (\ref{13})
for each triple of numbers $(Z,\lambda,N)$.
Those selected samples are presented in Tables \ref{T1}-\ref{T4} along with the corresponding parameters of the model WFs.

To solve the set of equations (\ref{8}), (\ref{9}) and $2(N-1)$ nonlinear equations of the form (\ref{13}) we apply, as the first step, the built-in function \textbf{NSolve[...]} of the Wolfram \emph{Mathematica}. The additional conditions (inequalities) $\textbf{b}_N>0$ are used.
The program \textbf{NSolve} generates all possible solutions. However, only one of them represents the nodeless solution which corresponds to the ground-state WF.
We have computed and presented the parameters of the model WFs for $3\leq N\leq6$.
It was mentioned above that the  \textbf{NSolve} is used only at the first step. The reason is that
this program works normally (with no problems) only for $N\leq 3$, that is for number of equations $2N\leq 6$. Even for $2N=6$ computer freezes for a few second capturing $100$ \%
of CPU time, and then normal operation is restored.
However, for $2N=8$ \emph{Mathematica} (through \textbf{NSolve}) takes all CPU time, and computer freezes for an indefinite time.   This is happened for any settings of \emph{Mathematica}, e.g., for any settings in "Parallel Kernel Configuration". We checked that this problem persists in different computers and for different version of \emph{Mathematica} (9, 10.3, 11.0, 12.1).
Therefore, to solve the relevant set of nonlinear equations for the number of exponentials $N>3$ we employed the built-in (\emph{Mathematica}) program \textbf{FindRoot[...]}.
Unlike \textbf{NSolve} this program generates only one solution (if it exists, of course) starting its search from some initial values $\textbf{C}_{N}^{(0)},\textbf{b}_{N}^{(0)}$ for which we take the values $\textbf{C}_{N-1},\textbf{b}_{N-1}$ of the corresponding calculation on the $N-1$ exponentials. The conditions of the positive exponents and the WF nodeless are certainly preserved.

To estimate the accuracy of the model WF we employ the following relation
\begin{equation}\label{14}
R_N=\int_0^\infty r|f_N(r)-\Phi(r,\lambda)|dr\left(\int_0^\infty r|\Phi(r,\lambda)|dr\right)^{-1}.
\end{equation}
Note that the function $r \Phi$ is more indicative than $\Phi$, at least, for the ground state.

\section{Results}\label{S2}

The two-exponential representations (excepting the case of $Z=1$) for the two-particle coalescences only (corresponding to the particular cases $\lambda=0$ and $\lambda=1$) were reported in Ref. \cite{LEZ5}.
In the current paper we calculate the parameters $\textbf{C}_N$ and $\textbf{b}_N$ of the model WFs, $f_N(r)\equiv f_N(r;\lambda)$ for the number of exponentials $3\leq N\leq 6$.
Our calculations are represented to different \emph{collinear} configurations including in particular the two-particle coalescences and the boundary case $\lambda=-1$.
The results are presented in Tables \ref{T1}-\ref{T4} together with the corresponding accuracy estimations $R_N$ and the sets $\left\{n_1,n_2,...,2(N-1)\right\}$ of integers included into the integrals (\ref{12}).
It is seen from all tables that the more exponentials generate the higher accuracy of the model WF.

\begin{figure}
\caption{The negative ion of hydrogen $\textrm{H}^-$($Z=1$):
\textbf{(\emph{a})} the WF, $\Phi(r,0)$ at the electron-nucleus coalescence (the \emph{collinear} configuration with $\lambda=0$) times $r$;
\textbf{(\emph{b})} the logarithmic estimates $\mathcal{L}_4^{(0}(r)$ and $\mathcal{L}_3^{(0}(r)$  of the difference (see Eq.(\ref{15})) between the model WF, $f_4(r)$ and the PLM WF (solid curve, blue online), and between the model WF, $f_3(r)$ and the PLM WF (dashed curve, red online), respectively.}
\includegraphics[width=6.0in]{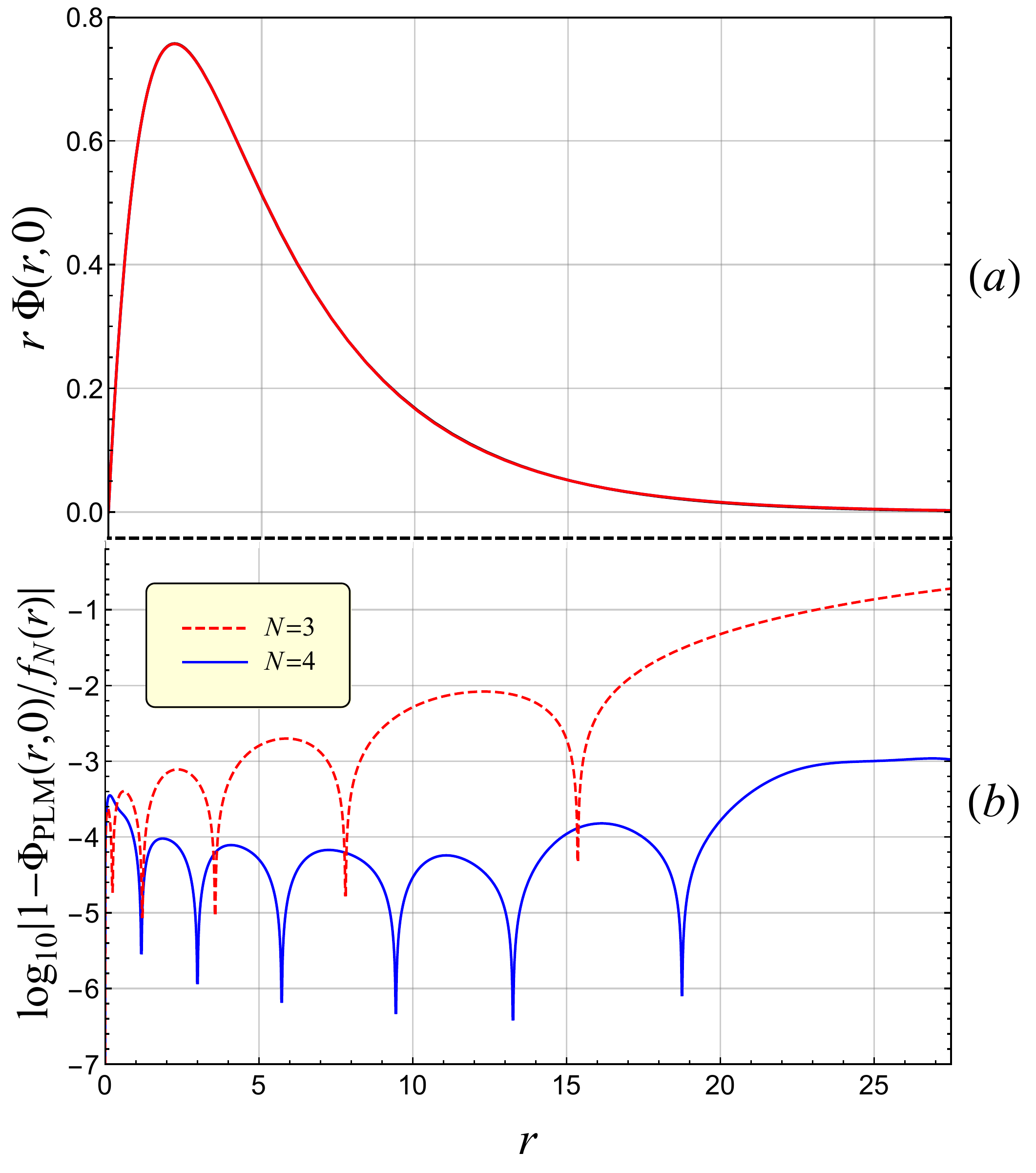}
\label{F1}
\end{figure}

\begin{figure}
\caption{The ground state of the positive ion of boron $\textrm{B}^{3+}$($Z=5$): \textbf{(\emph{a})}
 the WF, $\Phi(r,0)$ at the electron-nucleus coalescence (the \emph{collinear} configuration with $\lambda=0$) times $r$;
\textbf{(\emph{b})} the logarithmic estimates $\mathcal{L}_4^{(0}(r)$ and $\mathcal{L}_3^{(0}(r)$  of the difference between the model WF, $f_4(r)$ and the PLM WF (solid curve, blue online), and between the model WF, $f_3(r)$ and the PLM WF (dashed curve, red online), respectively.}
\includegraphics[width=6.0in]{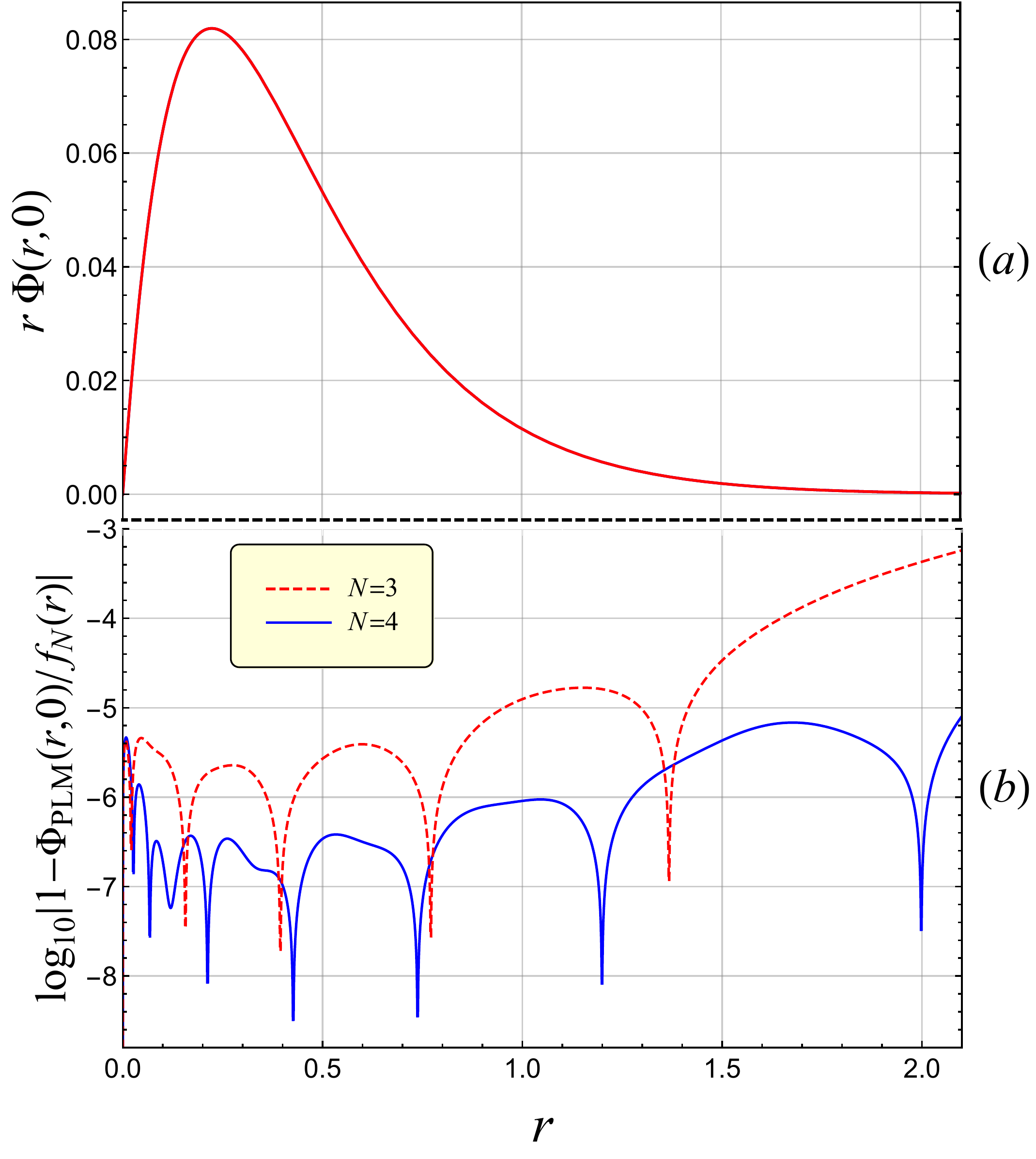}
\label{F2}
\end{figure}

One should notice that for $\lambda=0$, describing the case of the electron-nucleus coalescence, we were able to calculate the model WFs, $f_N(r)$ represented by three and four exponentials only ($N=3,4$). However, at least the case of $N=4$ shows very high accuracy, which is confirmed by the following.
Recall that the integral $R_N$ characterizes the general accuracy of $f_N(r)$.
In order to track changes in accuracy with distance $r$ we used the logarithmic function of the form
\begin{equation}\label{15}
\emph{L}_N^{(\lambda)}(r)=\log_{10}\left|1-f_N(r)/\Phi_{PLM}(r,\lambda)\right|.
\end{equation}
It is seen from Tables \ref{T1} and \ref{T2} that at least for $\lambda=0$ and given $N$ the minimal accuracy (represented by maximum $R_N$) is demonstrated by the negative ion $\textrm{H}^-$, whereas the maximum accuracy (represented by minimum $R_N$) is demonstrated by the positive ion $\textrm{B}^{3+}$.
The logarithmic functions $\emph{L}_N^{(0}(r)$ are shown in Figs.\ref{F1} and \ref{F2} for these two-electron ions with boundary (under consideration) nucleus charges $Z=1$ and $Z=5$.
It is seen that the deviations of the model WF from the PLM WF are practically uniform along the $r$-axis, and that one extra exponential improves accuracy by 1-2 (decimal) orders.
Regarding the accuracy of the model WF, $f_4(r)$ we would like to emphasize the following.
In Ref. \cite{LEZ3} (see figure $3b$ therein) it was displayed the logarithmic function $\mathcal{L}(r)$ of the form (\ref{15}) which describes the difference between the PLM WF and the CFHHM WF for the $\lambda=0$ \emph{collinear} configuration of the $\textrm{H}^-$ ion. The so called correlation function hyperspherical harmonic method (CFHHM) \cite{HM1, HM2} with the maximum  HH indices $K_m=128$ (1089 HH basis functions) was used for calculation of the fully (3-dimensional) WF  of the negative ion $\textrm{H}^-$.
Comparison of the logarithmic estimations $\emph{L}_4^{(0}(r)$ and the corresponding  $\mathcal{L}(r)$ shows that the model WF $f_4(r)$ is even more close to the PLM WF than the CFHHM WF for all values of $r$, which indicates on the \emph{extremely high accuracy} of the model WF (at least for $\lambda=0$ and $Z=1$) represented by \emph{four exponentials} only.
It is seen (see Fig. \ref{F2}) that the accuracy of the model WF $f_4(r)$ for $\textrm{B}^{3+}$ is higher about 2 decimal order than the 4-exponential WF for $\textrm{H}^-$. The logarithmic estimation $\mathcal{L}(r)$ for $\textrm{B}^{3+}$ is not presented in Ref. \cite{LEZ3}. However, the relevant calculations show that for this case ($\lambda=0$ and $Z=5$) the model WF $f_4(r)$ is more close to the PLM WF than the CFHHM WF, as well.

It was mentioned earlier that the behavior of the two-electron atomic WF near the nucleus is described by the Fock expansion (\ref{6}), which reduces to expansion (\ref{7}) for the \emph{collinear} arrangement of the particles.
The most compact model WFs represented by the sum of three or four exponentials were obtained for the case of the electron-nucleus coalescence corresponding to the \emph{collinear} parameter $\lambda=0$.
Tables \ref{T1} and \ref{T2} together with Figs.\ref{F1} and \ref{F2} demonstrate the high accuracy of those model WFs.
It should be emphasized that the accuracy of $f_4(r)$ for $\lambda=0$ is close to the accuracy of the variational PLM WF, $\Phi_{PLM}(r,0)$ for all $r>0$.
Furthermore, the relevant calculations show that the model WF $f_4(r)$ mentioned above is, in fact, more accurate than $\Phi_{PLM}(r,0)$ in the vicinity of nucleus ($r\rightarrow 0$).
We can argue this because the leading terms of the series expansion of $f_4(r)$ (for $\lambda=0$) are more close to the corresponding terms of the Fock expansion than the ones for $\Phi_{PLM}(r,0)$.
Actually, Eqs.(\ref{8}) and (\ref{9}) provide by definition the condition $f_4(0)=1$ and $f_4'(0)=-Z+1/2$, corresponding exactly to the Fock expansion.
Moreover, it is seen from Eq.(\ref{7b})) that for $\lambda=0$ the logarithmic term of the Fock expansion is annihilated because $\zeta_0=0$, and hence $F''(0)/2=\xi_0$, where we denoted $F(r)\equiv \Phi(r,0)$.
One should notice that $\lambda=0$ is, in fact, the single case of the \emph{collinear} arrangement when the explicit expression for the angular Fock coefficient $\xi_\lambda$ can be derived in the form \cite{LEZ3}, \cite{LEZ4}
\begin{equation}\label{16}
\xi_0=\frac{1-2E}{12}-Z\left(\frac{3-\ln 2}{6}\right)+\frac{1}{3}Z^2,
\end{equation}
where $E$ is the non-relativistic energy of the two-electron atom/ion under consideration.
It is seen from  Table \ref{T5} that (besides $f_4'(0)$) the values of $f_4''(0)/2$ is much closer to the theoretical values (\ref{16}) than $F_{PLM}''(0)/2$ for all $Z$.
These results confirm the above conclusion about the accuracy of the model WF
near the nucleus.

\section{Acknowledgments}\label{S3}

This work was supported by the PAZY Foundation, Israel.









\newpage

\begin{table}
\caption{The parameters of the model WFs $f_3(r)$.}
\begin{tabular}{|c|c|c|c|c|c|c|c|c|c|}
\hline
$\lambda$ &\footnotesize $Z$ & $b_1$ & $b_2$ & $b_3$ &\footnotesize $C_1$ & \footnotesize$C_2 $&\footnotesize $C_3$ &\scriptsize $n_1,n_2,n_3,n_4$  &\scriptsize$ 10^4 R_3$ \\
\hline
\hline\ft -1
   &\ft 1 &\sx 1.31221085126&\sx 2.30773912084&\sx 10.6895455708&\sx 1.21535118385 &\sx-0.203680351269 &\sx-0.0116708325859 &\sx 1,2,3,4&\sx 5.1\\
   &\ft 2 &\sx 3.30331779151&\sx 5.37300989029&\sx 23.5666840162&\sx 1.10218667377 &\sx-0.0971397085718&\sx-0.00504696520154&\sx 1,2,3,4&\sx 2.8\\
   &\ft 3 &\sx 5.29600197779&\sx 8.57050886387&\sx 37.1879160606&\sx 1.06342095849 &\sx-0.0603343891163&\sx-0.00308656937009&\sx 1,2,3,4&\sx 1.8\\
   &\ft 4 &\sx 7.29186771909&\sx 11.8104087791&\sx 51.8308243021&\sx 1.04554901504 &\sx-0.0433987968749&\sx-0.00215021816402&\sx 1,2,3,4&\sx 1.3\\
   &\ft 5 &\sx 9.28979602694&\sx 14.9934154050&\sx 64.5399905936&\sx 1.03580989481 &\sx-0.0340832362749&\sx-0.00172665853579&\sx 1,2,3,4&\sx 1.0\\
\hline\ft -0.5
   &\ft 1 &\sx 0.34947029202&\sx0.917929731034&\sx 3.34452782210&\sx0.00104484900062&\sx1.06791415059  &\sx-0.0689589995860 &\sx 2,4,6,8&\sx 26.1\\
   &\ft 2 &\sx 2.42281226939&\sx4.84214960756 &\sx 20.8946624090&\sx 1.05104853479 &\sx-0.0479768260983&\sx-0.00307170869253&\sx 2,3,4,5&\sx 0.82\\
   &\ft 3 &\sx 3.92179097813&\sx7.24103626268 &\sx 28.1658903414&\sx 1.03561982532 &\sx-0.0330601878282&\sx-0.00255963748729&\sx 2,3,4,5&\sx 0.38\\
   &\ft 4 &\sx 5.42080777652&\sx9.74980079227 &\sx 37.6087443483&\sx 1.02691550123 &\sx-0.0249667281386&\sx-0.00194877309539&\sx 2,3,4,5&\sx 0.31\\
   &\ft 5 &\sx 6.92015063432&\sx12.2629734705 &\sx 46.2827741201&\sx 1.02161443408 &\sx-0.0200074716239&\sx-0.00160696245286&\sx 2,3,4,5&\sx 0.25\\
\hline\ft 0
   &\ft 1 &\sx0.298919116361&\sx0.595029813767&\sx 7.19650438253&\sx 0.297067039246&\sx 0.704003189159 &\sx-0.00107022840522 &\sx 0,2,3,4&\sx 34.8\\
   &\ft 2 &\sx1.37487894240 &\sx1.79415111548 &\sx 6.51734657710&\sx 0.692940198275&\sx 0.307826353261 &\sx-0.000766551535934&\sx 2,3,4,5&\sx 0.16\\
   &\ft 3 &\sx2.38544848722 &\sx2.98500782211 &\sx 11.9948030706&\sx 0.806354227022&\sx 0.193817876564 &\sx-0.000172103585808&\sx 2,3,4,5&\sx 0.06\\
   &\ft 4 &\sx3.39003538779 &\sx4.18238632004 &\sx 44.1576144385&\sx 0.860366370158&\sx 0.139650495901 &\sx-0.0000168660592834&\sx 2,3,4,5&\sx 0.05\\
   &\ft 5 &\sx4.39300951665 &\sx5.38932092160 &\sx 77.5849424817&\sx 0.892332808812&\sx 0.107671063544 &\sx-3.87235633559$\times 10^{-6}$&\sx 2,3,4,5&\sx 0.05\\
\hline\ft 0.5
   &\ft 1 &\sx0.866272795833&\sx1.35400690862 &\sx 4.90509052234&\sx 0.445709671715&\sx 0.522361791814 &\sx 0.0319285364710 & \sx 2,3,4,5&\sx 13.7\\
   &\ft 2 &\sx2.43745271333 &\sx3.52128593757 &\sx 13.1425881206&\sx 0.790450763794&\sx 0.200669891111 &\sx 0.00887934509580& \sx 2,3,4,5&\sx 3.06\\
   &\ft 3 &\sx3.94735833664 &\sx5.62554439840 &\sx 21.0951038899&\sx 0.867392165040&\sx 0.127429856680 &\sx 0.00517797827994& \sx 2,3,4,5&\sx 1.72\\
   &\ft 4 &\sx5.45119270055 &\sx7.71384461880 &\sx 28.7740927992&\sx 0.902403168528&\sx 0.0938941380922&\sx 0.00370269337946& \sx 2,3,4,5&\sx 1.18\\
   &\ft 5 &\sx6.95360810088 &\sx9.81345881494 &\sx 36.9476130590&\sx 0.923088330095&\sx 0.0740947033493&\sx 0.00281696655580& \sx 2,3,4,5&\sx 0.91\\
\hline\ft 1
   &\ft 1 &\sx1.53675502654 &\sx2.33712280638 &\sx 8.25913664369&\sx 0.596515660760&\sx 0.379791535994 &\sx 0.0236928032457 & \sx 2,3,4,5&\sx 9.5\\
   &\ft 2 &\sx3.58172751587 &\sx5.24357995810 &\sx 19.2222658045&\sx 0.819528550521&\sx 0.172004590265 &\sx 0.00846685921408& \sx 2,3,4,5&\sx 2.85\\
   &\ft 3 &\sx5.58931707310 &\sx8.08649238609 &\sx 29.9307024541&\sx 0.881008315694&\sx 0.113793976078 &\sx 0.00519770822784& \sx 2,3,4,5&\sx 1.7\\
   &\ft 4 &\sx7.59237815227 &\sx10.9111334599 &\sx 40.2659747119&\sx 0.910825396831&\sx 0.0853703529884&\sx 0.00380425018111& \sx 2,3,4,5&\sx 1.19\\
   &\ft 5 &\sx9.59453605830 &\sx13.7549847587 &\sx 51.2908698042&\sx 0.929005400236&\sx 0.0680615517954&\sx 0.00293304796837& \sx 2,3,4,5&\sx 0.94\\
\hline
\end{tabular}
\label{T1}
\end{table}

\begin{table}
\caption{The parameters of the model WFs $f_4(r)$.}
\begin{tabular}{|c|c|c|c|c|c|c|c|}
\hline
$\lambda$ &\footnotesize $Z$  & $b_1$ &$b_2$&$b_3$ &$b_4$ &\scriptsize $n_1,n_2,n_3,n_4,n_5,n_6$& \\
 &  &\footnotesize $C_1$ &\footnotesize$C_2$ &\footnotesize$C_3$ &\footnotesize$C_4$ &  &\footnotesize$ 10^5 R_4$ \\
\hline
\hline\ft -1
 &\ft 1 &\ft 1.32020535772 & \ft 2.02749050880 & \ft 4.54443362311  & \ft 23.5929861083 &\ft 1,2,3,4,5,6 &\\
 &      &\sx 1.26227358934 & \sx-0.227201776295& \sx-0.0326343898304& \sx -0.00243742321816 & &\ft 2.7\\\cline{2-8}

 &\ft 2 &\ft 3.32040412399 & \ft 4.62849779909 & \ft 10.9532235110  & \ft 59.9204389330 &\ft 0,1,2,3,4,5&\\
 &      &\sx 1.14400570872 & \sx-0.128974625641& \sx-0.0142762170106& \sx -0.000754866072272 & &\ft 4.24\\\cline{2-8}

 &\ft 3 &\ft 5.31424074104 & \ft 7.36821061275 & \ft 17.2777611913  & \ft 91.9038190157 &\ft 0,1,2,3,4,5&\\
 &      &\sx 1.09045506589 &\sx-0.0810843733976&\sx-0.00889378923384& \sx-0.000476903253974 & &\ft 2.64\\\cline{2-8}

 &\ft 4 &\ft 7.30994394442 & \ft 10.1897566947 & \ft 24.4040098169  & \ft 148.641527280 &\ft 0,1,2,3,4,5&\\
 &      &\sx 1.06441904628 &\sx-0.0579854208721&\sx-0.00616816635722& \sx-0.000265459051415 & &\ft 2.18\\\cline{2-8}

 &\ft 5 &\ft 9.30868327139 & \ft 12.8750916787 & \ft 29.9556776119  & \ft 155.981269889 &\ft 0,1,2,3,4,5&\\
 &      &\sx 1.05138164659 &\sx-0.0460610340035&\sx-0.00504641308811& \sx-0.000274199500577 & &\ft 1.48\\

\hline\ft -0.5
 &\ft 1 &\ft 0.760415630622& \ft 0.980090816301& \ft 1.92593212146  & \ft 8.20874671468 &\ft 2,3,4,6,7,8&\\
 &      &\sx 0.170778873309& \sx 0.969455001129& \sx-0.130694209709 & \sx -0.00953966472852 & &\ft 11.0\\\cline{2-8}
 &\ft 2 &\ft 0.454124268500& \ft 2.42341297045 & \ft 4.76754499162  & \ft 19.5795924347 &\ft 2,3,4,5,6,7&\\
 &      &\sx{3.82312760898$ \times10^{-6}$}& \sx 1.05207275919& \sx-0.0486110885984 & \sx -0.00346549372265 & &\ft 8.6\\\cline{2-8}
 &\ft 3 &\ft 0.436440152659& \ft 3.92140440426 & \ft 7.30105365639  & \ft 28.9040922033 &\ft 2,3,4,5,6,7&\\
 &      &\sx{-7.40656697931$\times10^{-7}$}& \sx 1.03517571156& \sx-0.0327434647393 & \sx -0.00243150616122 & &\ft 8.3\\\cline{2-8}
 &\ft 4 &\ft 5.42161394245 & \ft 9.42798545298  & \ft 25.0471247671   & \ft 233.462987854 &\ft 1,2,3,4,5,6&\\
 &      &\sx 1.02782877628 & \sx-0.0249202692879& \sx-0.00283800779973& \sx{-7.04991926150$\times10^{-5}$} & &\ft 0.94\\\cline{2-8}
 &\ft 5 &\ft 6.92122844867 & \ft 11.7547534041  & \ft 29.3462817580   & \ft 174.466927481 &\ft 1,2,3,4,5,6&\\
 &      &\sx 1.02259238932 & \sx-0.0200065945557& \sx-0.00247182415422& \sx{-1.13970613965$\times10^{-4}$} & &\ft 0.55\\\cline{2-8}
\hline
\end{tabular}
\label{T2}
\end{table}

\begin{table}
\renewcommand\thetable{}
\caption{Continuation 1 of Table \ref{T2}.}
\begin{tabular}{|c|c|c|c|c|c|c|c|}
\hline
$\lambda$ &\footnotesize $Z$  & $b_1$ &$b_2$&$b_3$ &$b_4$ &\scriptsize $n_1,n_2,n_3,n_4,n_5,n_6$& \\
 &  &\footnotesize $C_1$ &\footnotesize$C_2$ &\footnotesize$C_3$ &\footnotesize$C_4$ &  &\footnotesize$ 10^5 R_4$ \\
\hline
\hline\ft 0
 &\ft 1 &\ft 0.262567575399& \ft 0.389274177978 & \ft 0.649724323343  & \ft 2.23082077059 &\ft 2,3,4,5,10,12&\\
 &      &\sx 0.124958793217& \sx 0.334742657170 & \sx 0.549255567787  & \sx-0.00895701817486 & &
 \ft 8.6\\\cline{2-8}
 &\ft 2 &\ft 0.863128362371& \ft 1.37696102972  & \ft 1.80094039452   & \ft 6.23029650467 &\ft 2,3,4,5,10,12&\\
 &    &\sx 0.000151625897771& \sx 0.700462341914 & \sx 0.300247696739  & \sx-0.000861664550097 & &
 \ft 0.85\\\cline{2-8}
 &\ft 3 &\ft 1.99107587529 & \ft 2.39179071871 & \ft 3.00926579184    & \ft 10.6422108618 &\ft 2,3,4,5,10,12&\\
 &    &\sx 0.00268122936659 & \sx 0.817445974254& \sx 0.180106425521   & \sx-0.000233629141750 & &
 \ft 0.16\\\cline{2-8}
 &\ft 4 &\ft 2.81017418133 & \ft 3.39598731473 & \ft 4.21438633241    & \ft 27.4015188295 &\ft 2,3,4,5,10,12&\\
 &    &\sx 0.00171603685935 & \sx 0.868965245468& \sx 0.129353924670   & \sx-3.52069972580$\times10^{-5}$ & &
 \ft 0.18\\\cline{2-8}
 &\ft 5 &\ft 4.38456769518 & \ft 5.14428012476 & \ft 6.14092966448    & \ft 16.8182459760 &\ft 2,3,4,5,10,12&\\
 &      &\sx 0.865405670253 & \sx 0.120564394928& \sx 0.0141051979746  & \sx-7.52631553724$\times10^{-5}$ & &
 \ft 0.047\\\cline{2-8}
\hline\ft 0.5
 &\ft 1 &\ft 0.813357864690& \ft 1.16216119635 & \ft 2.11035956249  & \ft 8.06302367772&\ft 2,3,4,5,6,7&\\
 &      &\sx 0.270203654224& \sx 0.601608930427& \sx 0.118017239524  & \sx 0.0101701758248 & &
 \ft 12.8\\\cline{2-8}
 &\ft 2 &\ft 2.39591108217 & \ft 3.04778032570 & \ft 5.77051889963  & \ft 23.1405644612&\ft 2,3,4,5,6,7&\\
 &      &\sx 0.666977476198& \sx 0.298527484633& \sx 0.0320148612526& \sx 0.00248017791627 & &
 \ft 2.7\\\cline{2-8}
 &\ft 3 &\ft 3.90776167717 & \ft 4.88664625816 & \ft 9.27871837715  & \ft 37.4614080792&\ft 2,3,4,5,6,7&\\
 &      &\sx 0.781821062131& \sx 0.198012796435& \sx 0.0187434533393& \sx 0.00142268809457& &
 \ft 1.5\\\cline{2-8}
 &\ft 4 &\ft 5.41235086929 & \ft 6.71460919468 & \ft 12.7195616951  & \ft 50.2714813824&\ft 2,3,4,5,6,7&\\
 &      &\sx 0.836936916554& \sx 0.148699140499& \sx 0.0133242070845& \sx 0.00103973586269& &
 \ft 1.0\\\cline{2-8}
 &\ft 5 &\ft 6.91531456746 & \ft 8.55334168563 & \ft 16.2935883030  & \ft 66.2413275150&\ft 2,3,4,5,6,7&\\
 &      &\sx 0.870584918595& \sx 0.118467258140& \sx 0.0101878055295& \sx 0.000760017736373& &
 \ft 0.81\\\cline{2-8}
\hline
\end{tabular}
\label{T2a}
\end{table}

\begin{table}
\renewcommand\thetable{}
\caption{Continuation 2 of Table \ref{T2}.}
\begin{tabular}{|c|c|c|c|c|c|c|c|}
\hline
$\lambda$ &\footnotesize $Z$  & $b_1$ &$b_2$&$b_3$ &$b_4$ &\scriptsize $n_1,n_2,n_3,n_4,n_5,n_6$& \\
 &  &\footnotesize $C_1$ &\footnotesize$C_2$ &\footnotesize$C_3$ &\footnotesize$C_4$ &  &\footnotesize$ 10^5 R_4$ \\
\hline
 \hline \footnotesize{$~1~$}
 &\ft 1 &\ft 1.45895844248 & \ft 1.94229863142 & \ft 3.64144369441  & \ft 14.6402400267&\ft 2,3,4,5,6,7&\\
 &      &\sx 0.374761964378& \sx 0.526163652625& \sx 0.0926648508651& \sx 0.00640953213255 & &
 \ft 11.9\\\cline{2-8}
 &\ft 2 &\ft 3.52660868338 & \ft 4.46313533469 & \ft 8.41366993352  & \ft 33.9863982377&\ft 2,3,4,5,6,7&\\
 &      &\sx 0.702939199818& \sx 0.262807685971& \sx 0.0319120324665& \sx 0.00234108174421 & &
 \ft 2.9\\\cline{2-8}
 &\ft 3 &\ft 5.53691125164 & \ft 6.92690040645 & \ft 13.1173116907  & \ft 53.3038611455&\ft 2,3,4,5,6,7&\\
 &      &\sx 0.800429839233& \sx 0.178804526892& \sx 0.0193437507870& \sx 0.00142188308737 & &
 \ft 1.7\\\cline{2-8}
 &\ft 4 &\ft 7.54061400125 & \ft 9.36808709974 & \ft 17.6989833742  & \ft 70.3315968499&\ft 2,3,4,5,6,7&\\
 &      &\sx 0.848381505855& \sx 0.136508216332& \sx 0.0140382215614& \sx 0.00107205625181& &
 \ft 1.1\\\cline{2-8}
 &\ft 5 &\ft 9.54371206725 & \ft 11.8356042739 & \ft 22.5015053896  & \ft 91.8749584959&\ft 2,3,4,5,6,7&\\
 &      &\sx 0.878994353474& \sx 0.109401348385& \sx 0.0108088103065& \sx 0.000795487835141& &
 \ft 0.91\\\cline{2-8}
\hline
\end{tabular}
\label{T2b}
\end{table}

\setcounter{table}{2}
\begin{table}
\caption{The parameters of the model WFs $f_5(r)$.}
\begin{tabular}{|c|c|c|c|c|c|c|c|c|}
\hline
$\lambda$ &\footnotesize $Z$  & $b_1$ &$b_2$&$b_3$ &$b_4$ &$b_5$&\scriptsize $n_1,n_2,n_3,n_4,$& \\
 &  &\footnotesize $C_1$ &\footnotesize$C_2$ &\footnotesize$C_3$ &\footnotesize$C_4$&\footnotesize$C_5$ &\scriptsize $n_5,n_6,n_7,n_8$  &\footnotesize$ 10^6 R_5$ \\
\hline
\hline\ft -1
 &\ft 2 &\ft 3.33403942234 & \ft 4.15321766999 & \ft 7.22771669209  & \ft 18.0394750352   &\ft 98.9124103713 &\ft 0,1,2,3,4,5,6,7&\\
 &      &\sx 1.19779177816 & \sx-0.165381648981& \sx-0.0277431203764& \sx-0.00439609964953&\sx-0.000270909151645& &\ft 4.5\\\cline{2-9}
 &\ft 3 &\ft 5.32933696681 & \ft 6.59886648653 & \ft 11.4228762079  & \ft 28.2096051133   &\ft 145.866483798 &\ft 0,1,2,3,4,5,6,7&\\
 &      &\sx 1.12638430126 & \sx-0.105910712803& \sx-0.0175600965385& \sx-0.00273317456436&\sx-0.000180317352143& &\ft 2.8\\\cline{2-9}
 &\ft 4 &\ft 7.32573160412 & \ft 9.10153474687 & \ft 15.9377755826  & \ft 40.8969472027   &\ft 290.857815110 &\ft 0,1,2,3,4,5,6,7&\\
 &      &\sx 1.09052624934 & \sx-0.076177131758& \sx-0.0124353779117& \sx-0.00183726599787&\sx-7.64736726593$\times10^{-5}$& &\ft 2.4\\\cline{2-9}
 &\ft 5 &\ft 9.32452700647 & \ft 11.5219291480 & \ft 19.8392221432  & \ft 48.4980928929   &\ft 239.735778051 &\ft 0,1,2,3,4,5,6,7&\\
 &      &\sx 1.07219627060 & \sx-0.060471688997& \sx-0.0100638089132& \sx-0.00155216378212&\sx-0.000108608909614& &\ft 1.5\\\cline{2-9}
 \hline\ft -0.5
 &\ft 1 &\ft 0.782723673384  & \ft 1.00579499131 & \ft 1.67516053163  & \ft 3.84347571684   &\ft 20.7036985208 &\ft 1,2,3,4,6,7,8,9&\\
 &      &\sx 0.247875527976  & \sx 0.932587577289& \sx-0.156445025546 & \sx-0.0223795177635 &\sx-0.00163856195617& &\ft 12.3\\\cline{2-9}
 &\ft 2 &\ft 2.01254813456   & \ft 2.43474345293 & \ft 4.24085899493  & \ft 9.91825838265   &\ft 49.9226239551 &\ft 0,2,3,4,5,6,7,9&\\
 &      &\sx0.0111225319077  & \sx 1.05138001690 & \sx-0.0543797090181& \sx-0.00759679366215&\sx-0.000526046127036& &\ft 6.3\\\cline{2-9}
 &\ft 3 &\ft 2.66771135582   & \ft 3.92527807057 & \ft 6.68858179039  & \ft 15.1815880350   &\ft 69.4634730215 &\ft 1,2,3,4,5,6,7,8&\\
 &      &\sx0.000323581113314& \sx 1.03908286278 & \sx-0.0341431378917& \sx-0.00487131736096&\sx-0.000391988635926& &\ft 3.6\\\cline{2-9}
 &\ft 4 &\ft 5.02658424937   & \ft 5.43330301186 & \ft 8.82243334116  & \ft 19.5615991527   &\ft 96.7309045026 &\ft 1,2,3,4,5,6,7,8&\\
 &      &\sx 0.0176544037925 & \sx 1.01381753461 & \sx-0.0270777601306& \sx-0.00410558781973&\sx-0.000288590450437& &\ft 3.6\\\cline{2-9}
 &\ft 5 &\ft 6.76854109699   & \ft 6.93768261467 & \ft 11.4098098294  & \ft 26.4501588397   &\ft 130.894355406 &\ft 0,1,2,3,4,5,6,7&\\
 &      &\sx 0.08493484927950& \sx 0.939048336705& \sx-0.0209035403321& \sx-0.00289062795813&\sx-0.000189017693978& &\ft 3.7\\\cline{2-9}
\hline
\end{tabular}
\label{T3}
\end{table}

\begin{table}
\renewcommand\thetable{}
\caption{Continuation of Table \ref{T3}.}
\begin{tabular}{|c|c|c|c|c|c|c|c|c|}
\hline
$\lambda$ &\footnotesize $Z$  & $b_1$ &$b_2$&$b_3$ &$b_4$ &$b_5$&\scriptsize $n_1,n_2,n_3,n_4,$& \\
 &  &\footnotesize $C_1$ &\footnotesize$C_2$ &\footnotesize$C_3$ &\footnotesize$C_4$&\footnotesize$C_5$ &\scriptsize $n_5,n_6,n_7,n_8$  &\footnotesize$ 10^6 R_5$ \\
\hline
 \hline\ft 0.5
 &\ft 1 &\ft 0.790902643508  & \ft 1.06035478911 & \ft 1.50532224818  & \ft 3.03736727252   &\ft 11.1410180677 &\ft 2,3,4,5,6,7,8,9&\\
 &      &\sx 0.196607406672  & \sx 0.508558045232& \sx 0.247655549880 & \sx 0.0423479201908 &\sx 0.00483107802464& &\ft 20.9\\\cline{2-9}
 &\ft 2 &\ft 2.37639295406   & \ft 2.86725870274 & \ft 4.29974756696  & \ft 8.77965927735   &\ft 35.0370610367 &\ft 2,3,4,5,6,7,8,9&\\
 &      &\sx 0.590329529453  & \sx 0.341324441817& \sx 0.0576037324822& \sx 0.00973365189967&\sx 0.00100864434805& &\ft 3.8\\\cline{2-9}
 &\ft 3 &\ft 3.88798021172   & \ft 4.60274029470 & \ft 6.92722664338  & \ft 14.1907934142   &\ft 57.3767081273 &\ft 2,3,4,5,6,7,8,9&\\
 &      &\sx 0.721150604506  & \sx 0.238651951852& \sx 0.0340403271804& \sx 0.00558868261345&\sx 0.000568433848049& &\ft 2.1\\\cline{2-9}
 &\ft 4 &\ft 5.39378534978   & \ft 6.35210931219 & \ft 9.65638723913  & \ft 19.9037177229   &\ft 76.9969095748 &\ft 2,3,4,5,6,7,8,9&\\
 &      &\sx 0.791446509718  & \sx 0.181080013887& \sx 0.0233325486312& \sx 0.00373563625064&\sx 0.000405291512930& &\ft 1.2\\\cline{2-9}
 &\ft 5 &\ft 6.89631641077   & \ft 8.08529427507 & \ft 12.3139186919  & \ft 25.6710836429   &\ft 107.258963467 &\ft 2,3,4,5,6,7,8,9&\\
 &      &\sx 0.831962265527  & \sx 0.146815487037& \sx 0.0180923679551& \sx 0.00285602492664&\sx 0.000273854553497& &\ft 1.1\\\cline{2-9}
 \hline\ft 1
 &\ft 1 &\ft 1.40532919228   & \ft 1.76542107311 & \ft 2.67094532497  & \ft 5.73653600501   &\ft 28.8596269922 &\ft 2,3,4,5,6,7,8,9&\\
 &      &\sx 0.228872138883  & \sx 0.564008891655& \sx 0.177278010589 & \sx 0.0281992012650 &\sx 0.00164175760899& &\ft 26.2\\\cline{2-9}
 &\ft 2 &\ft 3.49162225456   & \ft 4.12954660811 & \ft 6.24917362659  & \ft 13.0406927609   &\ft 55.6660835808 &\ft 2,3,4,5,6,7,8,9&\\
 &      &\sx 0.600886634251  & \sx 0.328953425818& \sx 0.0598766127109& \sx 0.00945702883157&\sx 0.000826298388269& &\ft 5.2\\\cline{2-9}
 &\ft 3 &\ft 5.50808155281   & \ft 6.48741655765 & \ft 10.0193133315  & \ft 21.4634677872   &\ft 95.2150755643 &\ft 2,3,4,5,6,7,8,9&\\
 &      &\sx 0.736257943800  & \sx 0.224466915903& \sx 0.0337303661233& \sx 0.00511826651080&\sx 0.000426507663627& &\ft 2.8\\\cline{2-9}
 &\ft 4 &\ft 7.51732319016   & \ft 8.88501166054 & \ft 14.1919981920  & \ft 32.7751210282   &\ft 162.408084363 &\ft 1,2,3,4,5,6,7,8&\\
 &      &\sx 0.807735699959  & \sx 0.166935972380& \sx 0.0220982375951& \sx 0.00303510193517&\sx 0.000194988130580& &\ft 2.5\\\cline{2-9}
 &\ft 5 &\ft 9.52058905892   & \ft 11.2258539229 & \ft 17.9584770013  & \ft 41.6932712805   &\ft 236.717424813 &\ft 1,2,3,4,5,6,7,8&\\
 &      &\sx 0.845567446616  & \sx 0.134819493378& \sx 0.0171241354063& \sx 0.00236103746563&\sx 0.000127887134412& &\ft 2.0\\\cline{2-9}
\hline
\end{tabular}
\label{T3a}
\end{table}

\setcounter{table}{3}

\begin{table}
\caption{The parameters of the model WFs $f_6(r)$ for the negative ion of hydrogen ($Z=1$).}
\begin{tabular}{|c|c|c|c|c|c|c|c|c|}
\hline
$\lambda$ & $b_1$ &$b_2$&$b_3$ &$b_4$ &$b_5$&$b_6$&\scriptsize $n_1,n_2,$& \\
 &  \footnotesize $C_1$ &\footnotesize$C_2$ &\footnotesize$C_3$ &\footnotesize$C_4$&\footnotesize$C_5$&\footnotesize$C_6$&\scriptsize $...,n_{10}$ &\footnotesize$ 10^6 R_6$ \\
\hline
\hline\ft -1
 &\sx 0.616882428065      & \sx 1.32144483672 & \sx 1.98116133623  & \sx 3.89714002309   &\sx 10.7703678294   &\sx 117.105539565     &\ft 0,1,2,&\\
       &\sx 0.0000220356978505  & \sx 1.27099937960 & \sx-0.227846680046 & \sx-0.0372255443263 &\sx-0.00577032259649&\sx-0.000178868331372 &\ft ...,9 &\ft 7.7\\
\hline\ft -0.5
  &\sx 0.784443648962      & \sx 1.00933647226 & \sx 1.62815973825  & \sx 3.28191702383   &\sx 9.07710968927   &\sx 175.255656318     &\ft 0,1,2,&\\
      &\sx 0.255668095458      & \sx 0.933467043932& \sx-0.159437110775 & \sx-0.0255522308928 &\sx-0.00407564245776&\sx-0.0000701552652768&\ft ...,9 &\ft 3.5\\
\hline\ft 0.5
  &\sx 0.787145306516      & \sx 1.03714737333 & \sx 1.42318385820  & \sx 2.66484832743   &\sx 7.69976806889   &\sx 534.086213753     &\ft 0,1,2,&\\
      &\sx 0.183792703068      & \sx 0.466750597846& \sx 0.288402624695 & \sx 0.0530568221124 &\sx 0.00798238877269&\sx 0.0000148635064310&\ft ...,9 &\ft 7.2\\
\hline\ft 1
  &\sx 1.36357313996      & \sx 1.649819114196 & \sx 2.20488118195  & \sx 3.61727560292   &\sx 8.21954358047   &\sx 125.741386557     &\ft 2,3,4,&\\
      &\sx 0.134755301707     & \sx 0.5137129189258& \sx 0.272937433228 & \sx 0.0657008289814 &\sx 0.0126953713973 &\sx 0.000198145760468 &\ft ...,11 &\ft 4.8\\
 \hline
\end{tabular}
\label{T4}
\end{table}

\begin{table}
\caption{The first and second derivatives of the \emph{collinear} WF with $\lambda=0$ at the nucleus.
The PLM WF, $F(r)\equiv \Phi_{PLM}(r,0)$ at the electron-nucleus coalescence is introduced.}
\begin{tabular}{|c|c|c|c|c|c|}
\hline
$Z$& $F'(0)$&$-Z+1/2$ & $F''(0)/2$ &$f_4''(0)/2$&$\xi_0$ \\
\hline
\hline\ 1 & -0.506379  &-0.5  &0.169101   & 0.123314 &0.12015   \\
\hline\ 2 & -1.50228   &-1.5  &1.20558    & 1.13429  &1.13167   \\
\hline\ 3 & -2.50175   &-2.5  &3.24348    & 3.14574  &3.14323   \\
\hline\ 4 & -3.50323   &-3.5  &6.38221    & 6.15306  &6.15469   \\
\hline\ 5 & -4.50140   &-4.5  &10.3165    & 10.1691  &10.1661   \\
 \hline
\end{tabular}
\label{T5}
\end{table}

\bibliography{mybib}
\end{document}